\def\kx{{{\hat{k}}_x}}
\def\ky{{{\hat{k}}_y}}
\def\kz{{{\hat{k}}_z}}

\def\eqi{\begin{equation}}
\def\eqf{\end{equation}}
\def\eqia{\begin{eqnarray}}
\def\eqfa{\end{eqnarray}}
\def\Om{\mathit{\Omega}}
\def\rp#1#2{{#1\over#2}}
\def\lb#1{\label{#1}}
\def\kap{\bds{\hat{k}}}

\def\bds#1{\boldsymbol{#1}}

\def\cO{\cos\Om}
\def\sO{\sin\Om}

\def\cI{\cos I}
\def\sI{\sin I}

\def\ton#1{\left(#1\right)}
\def\qua#1{\left[#1\right]}
\def\grf#1{\left\{#1\right\}}

\documentclass{ws-procs975x65}
\usepackage{wasysym}
\usepackage[super,compress]{cite}
\def\beq{\begin{equation}}
\def\eeq{\end{equation}}

\begin{document}

\title{The Solar Lense-Thirring effect: \\ perspectives for a future measurement}
\author{Lorenzo Iorio$^*$}

\address{Ministero dell' Istruzione, dell' Universit\`{a} e della Ricerca-Istruzione,\\
Viale Unit\`{a} di Italia 68, Bari,  70125, Italy\\
$^*$E-mail: lorenzo.iorio@libero.it\\
http://digilander.libero.it/lorri/homepage\_of\_lorenzo\_iorio.htm}

\begin{abstract}
The  predicted Lense-Thirring perihelion precession of Mercury induced by the Sun's angular momentum through its general relativistic gravitomagnetic field amounts to 2 milliarcseconds per century. It turned out to be compatible with the latest experimental determinations of the supplementary perihelion precession of Mercury with the INPOP15a ephemerides, whose accuracy level has nowadays reached the magnitude of the predicted relativistic effect itself thanks to the analysis of some years of tracking data of the MESSENGER spacecraft, which orbited Mercury from 2011 to 2015. A dedicated analysis of three years of MESSENGER data with the DE ephemerides allowed for a $25\%$ determination of the Sun's angular momentum by means of the Lense-Thirring effect, which turned out to be highly correlated with the signature due to the Solar quadrupole mass moment $J_2^{\odot}$.
\end{abstract}

\keywords{Experimental studies of gravity; Experimental tests
of gravitational theories; General relativity and gravitation; Ephemerides, almanacs, and calendars}

\bodymatter

\section{The Lense-Thirring Pericenter Precession}
In its weak-field and slow-motion approximation, the General Theory of Relativity (GTR) predicts that the orbital motion of a test body around a slowly rotating mass is affected by the so-called gravitomagnetic component of the central object's gravitational field due to the proper angular momentum $\bds S$ of the latter. As a result, the Lense-Thirring (LT) secular precessions of some of the orbital elements of the particle occur\cite{LT18}. In particular, the  LT precession $\dot\varpi_{\rm LT}$ of the longitude of periapsis $\varpi\doteq\Om + \omega$, where $\Om,~\omega$ are the longitude of the ascending node and the argument of periapsis, respectively, is\cite{Iorio012}
\begin{align}
\dot\varpi_{\rm LT} \nonumber &= -\rp{2GS}{c^2 a^3\ton{1-e^2}^{3/2}}\grf{2\qua{\kz\cI +\sI\ton{\kx\sO -\ky\cO}  } -\right.\\ \nonumber \\
&- \left. \qua{\kz\sI +\cI\ton{\ky\cO -\kx\sO} }\tan\ton{\rp{I}{2}}}\lb{oLT}.
\end{align}
 In Eq.~\ref{oLT}, $G,~c$ are the Newtonian constant of gravitation and the speed of light, respectively, $\kx,~\ky,~\kz$ are the components of the body's spin unit vector $\kap$ in the coordinate system adopted, while $a,~e,~I$ are the particle's semimajor axis, eccentricity and inclination of the orbital plane to the reference $\grf{x,~y}$ plane adopted, respectively. Eq \ref{oLT} is the sum of the LT precessions\cite{Iorio012} of the node and the argument of periapsis for an arbitrary orientation of $\bds S=S\kap$.

So far, more or less reliable and accurate attempts to detect the  LT effect have been made, or will be performed, with spacecraft orbiting some planets of the Solar System like the Earth, Mars and Jupiter; see, e.g., Refs.~\citen{Ciuf011, Iorioetal011, Renz013} and references therein. the Gravity Probe B (GP-B) mission recently measured another gravitomagnetic effect, i.e. the Pugh-Schiff gyroscope precession, in the field of the Earth, although with a lower accuracy than expected\cite{GPB}. The main difficulties in accurately measuring the LT effect reside in its generally small magnitude and in the relatively huge competing effects of classical origin whose modeling may not always be accurate enough to allow for a reliable subtraction from the data processed.
\section{The Sun's Lense-Thirring effect for Mercury}
In the case of the Sun, it is
\eqi\kx^{\odot} = 0.122,~\ky^{\odot} = -0.423,~\kz^{\odot} = 0.897\eqf
in a coordinate system having the mean Earth's equator at the J2000.0 epoch as reference $\grf{x,~y}$ plane. Since, according to helioseismology\cite{Pijp98}, the size of the Solar angular momentum is
\eqi S_{\odot} = 1.92\times 10^{41}~\textrm{kg}~\textrm{m}^2~\textrm{s}^{-1},\eqf
the expected LT perihelion precession of Mercury is at the milliarcseconds~per~century (mas~cty$^{-1}$) level; indeed, according to Eq.~\ref{oLT}, it amounts to
\eqi\dot\varpi_{\rm LT}^{\mercury} = -2.0~\textrm{mas}~\textrm{cty}^{-1}.\lb{MerLT}\eqf

As shown in Ref.~\refcite{Iorioetal011} and references therein,  from some years now it has been noted that steady advances in the orbit determination of Mercury may allow to measure the tiny Hermean perihelion precession of Eq.~\ref{MerLT} in a not too far future. To this respect, the latest results in the field of planetary ephemerides seem somewhat encouraging since they show that the accuracy reached nowadays in Mercury's orbit determination, after the analysis of the tracking data of the MESSENGER spacecraft from 2011 to 2014 with the INPOP15a ephemerides\cite{Fie015}, is at the level of the GTR prediction of Eq.~\ref{MerLT} itself. Indeed, the average of two determinations of the supplementary perihelion precession $\Delta\dot\varpi$ of Mercury with different criteria reads\cite{Fie016}
\eqi\left\langle\Delta\dot\varpi_{\rm exp}^{\mercury}\right\rangle = 0.0\pm 2.075~\textrm{mas}~\textrm{cty}^{-1}.\lb{Fie}\eqf
 The parameter $\Delta\dot\varpi$ represents the experimentally allowed range for any possible departure of a planet's perihelion precession from its standard value as predicted by GTR and Newtonian dynamics, (almost) fully modeled in the data reduction softwares of the ephemerides to the best of our current knowledge of them. Since Sun's gravitomagnetism was not explicitly included in the INPOP15a ephemerides, Eq.~\ref{MerLT} should entirely account for it, at least in principle, along with the mismodeled components of the precession due to other classical dynamical effects such as the Solar quadrupole mass moment $J_2^{\odot}$, the ring of minor asteroids, etc. The result of Eq.~\ref{Fie} shows that, at least, the GTR prediction of Eq.~\ref{MerLT} is compatible with the most recent orbit determinations of Mercury, although it is not yet possible to measure it.
 
A different, dedicated approach was followed by the authors of Ref.~\refcite{Park015} who processed three years of ranging data to MESSENGER by explicitly modeling the gravitomagnetic field of the Sun, and attempted to simultaneously estimate it along with other relevant physical parameters of the Sun and the spacecraft itself. They found that the Lense-Thirring effect is highly correlated with that due to $J_2^{\odot}$, and a determination of the relativistic signature of interest with an accuracy of about $25\%$ is possible.

The analysis of the full tracking data record of MESSENGER, which ended its mission by crashing onto the Hermean surface on April 30, 2015, and the forthcoming BepiColombo mission to Mercury, scheduled to be launched in 2017, should further improve the perspectives of measuring this elusive general relativistic effect.

\end{document}